\documentclass[aps,prl,floatfix,twocolumn,tightenlines]{revtex4}
\pdfoutput=1
\usepackage[pdftex]{hyperref}
\usepackage{amsmath}
\usepackage{latexsym}
\usepackage{graphicx}
\usepackage{color}
\usepackage{natbib}

\begin{document}

\title{Classical analogues of two-photon quantum interference}
\author{R. Kaltenbaek}
\affiliation{Institute for Quantum Computing and Department of Physics \&
Astronomy, University of Waterloo, Waterloo, Canada, N2L 3G1}
\author{J. Lavoie}
\affiliation{Institute for Quantum Computing and Department of Physics \&
Astronomy, University of Waterloo, Waterloo, Canada, N2L 3G1}
\author{K.J. Resch}
\email[Corresponding author: ]{kresch@iqc.ca}
\affiliation{Institute for Quantum Computing and Department of Physics \&
Astronomy, University of Waterloo, Waterloo, Canada, N2L 3G1}

\begin{abstract}
\noindent Chirped-pulse interferometry (CPI) captures the
metrological advantages of quantum Hong-Ou-Mandel (HOM)
interferometry in a completely classical system.  Modified HOM
interferometers are the basis for a number of seminal quantum
interference effects. Here, the corresponding modifications to CPI
allow for the first observation of classical analogues to the HOM
peak and quantum beating. They also allow a new classical technique
for generating phase super-resolution exhibiting a coherence length
dramatically longer than that of the laser light, analogous to
increased two-photon coherence lengths in entangled states.
\end{abstract}
\maketitle

Quantum-optics experiments demonstrated a wide range of interference
phenomena that had never before been seen in classical systems.
Prominent examples include: automatic dispersion and aberration
cancellation~\cite{Steinberg1992a,Franson1992a,Bonato2008a},
phase-insensitive interference~\cite{Hong1987a}, nonlocal
interference~\cite{Franson1989a, Ou1990b}, ghost
imaging~\cite{Pittman1995a} \& ghost
diffraction~\cite{Strekalov1995a}, phase
super-resolution~\cite{Rarity1990a,Mitchell2004a,Walther2004a}, and
phase
super-sensitivity~\cite{Yurke1986a,Lee2002a,Giovannetti2004a,Higgins2007a}.
Some of these phenomena form the basis for applications in quantum
computing and metrology that promise to outperform their classical
counterparts in terms of speed and precision, respectively.
Recently, ghost imaging~\cite{Bennink2002a, Ferri2005a}, automatic
dispersion
cancellation~\cite{Resch2007b,Kaltenbaek2008a,Lavoie2009a}, phase
super-resolution \cite{Resch2007a}, and phase insensitive
interference~\cite{Kaltenbaek2008a} have been observed in classical
optical systems exploiting correlation, but not entanglement.
Chirped-pulse interferometry (CPI) \cite{Kaltenbaek2008a} is a new,
completely classical technique producing the same interferogram as a
Hong-Ou-Mandel (HOM) interferometer \cite{Hong1987a} based on
frequency-entangled photon pairs, but with vastly higher signal. It
has been shown that modifications to the HOM interferometer can
produce a wide array of quantum interference effects such as the HOM
peak \cite{Mattle1996a}, quantum beating \cite{Ou1988a,Legero2004a},
and phase super-resolution~\cite{Rarity1990a}.  In the present work,
we show how similar modifications to CPI can produce the analogous
interferometric signatures with only classical resources. Thus we
rule out the HOM peak and quantum beating signatures as uniquely
quantum and demonstrate phase super-resolution in a classical
context with important differences from previous work
\cite{Resch2007a}.

\begin{figure}
\begin{center}
\label{setup}
\includegraphics[width=1 \columnwidth]{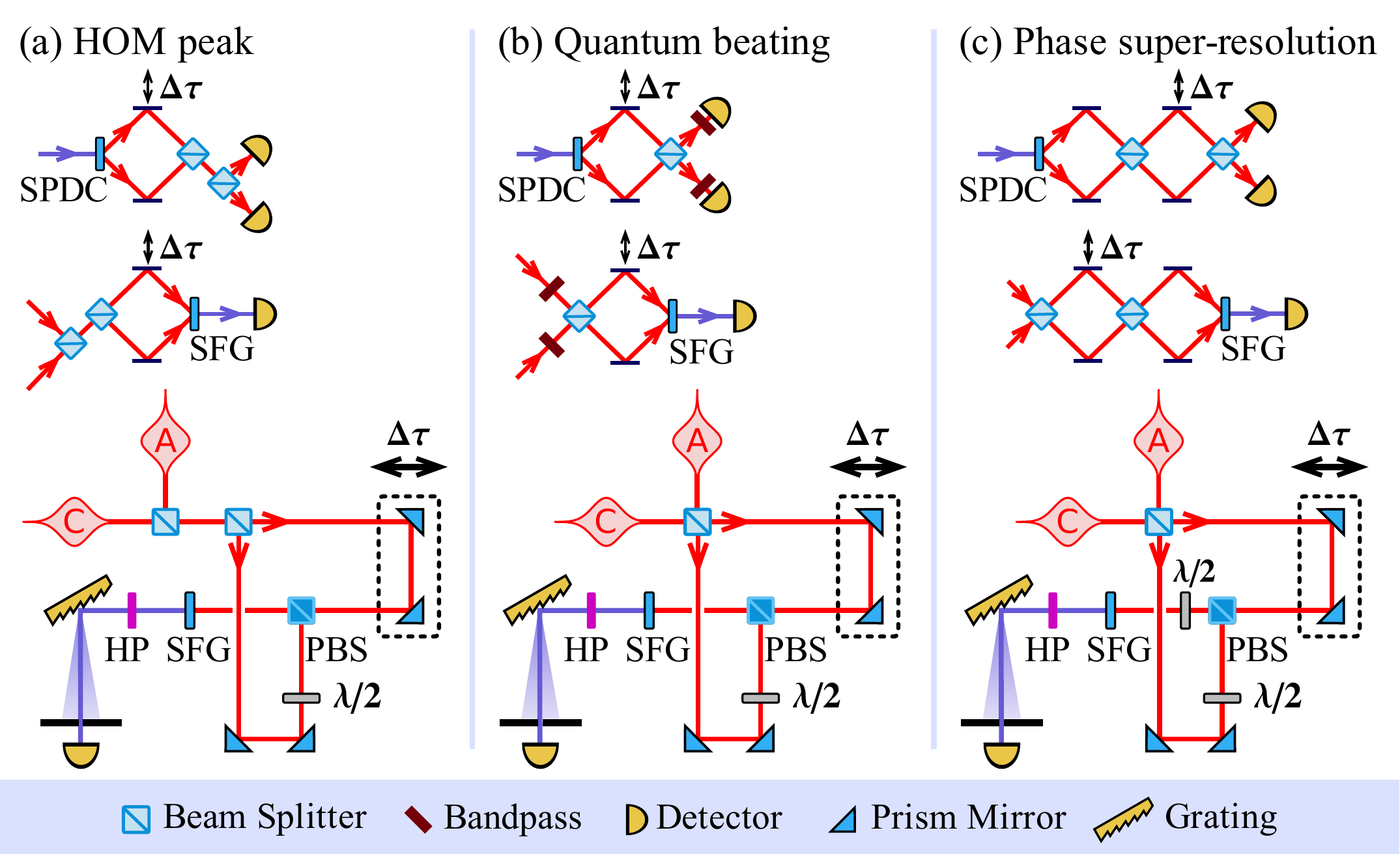}
\caption{Two-photon interferometers and their time-reversed
analogues. The three top figures show schematic of the quantum
interferometers used to observe a) the HOM peak, b) quantum beating
(note the inclusion of two filters with different bandpasses), and
c) two-photon phase super-resolution. All of the interferometers
rely on light created from spontaneous parametric down-conversion
(SPDC) in a nonlinear crystal. The interferograms correspond to the
number of coincidence counts registered at a pair of photon counting
detectors as a function of the path delay, $\Delta\tau$. The middle
row of figures depict the time-reversed version of each quantum
interferometer based on the recently described chirped-pulse
interferometry~\cite{Kaltenbaek2008a}. These time-reversed
interferometers were implemented as shown in the bottom row of
figures. Chirped (C) and and anti-chirped (A) laser pulses with
matched chirp rates are combined at the inputs of the
interferometers. The light passes along the two arms of the
interferometer, is recombined at a polarizing beam splitter (PBS),
and is focused on a nonlinear crystal. High-pass filters (HP) remove
the fundamental from the resulting sum-frequency generation (SFG). A
narrow band of frequencies is filtered, via gratings and a slit, and
detected via an amplified Si photodiode. \label{schematics}}
\end{center}
\end{figure}

Hong-Ou-Mandel interference~\cite{Hong1987a} is ubiquitous in
optical quantum information processing, underlying such effects as
quantum teleportation~\cite{Bennett1993a, Bouwmeester1997a} and
linear-optics quantum computation~\cite{Knill2001a}. It occurs when
two photons are coherently combined on a beamsplitter, and manifests
as a dip in the coincidence rate of two detectors. A typical HOM
interferometer, apart from the bandpass filters, is depicted in
Fig.~\ref{schematics}b)~(upper). HOM interference with
frequency-entangled photons exhibits automatic dispersion
cancellation, phase insensitivity and robustness against loss,
rendering it a promising tool for quantum metrology and 
imaging~\cite{Steinberg1992a,Abouraddy2002a,Nasr2003a}. We have
recently demonstrated chirped-pulse interferometry
~\cite{Kaltenbaek2008a}, a completely classical technique that exhibits all 
of these important features of HOM interference.
This classical approach can be viewed as a time-reversed version of
the HOM interferometer~\cite{Resch2007a}, see Fig.~1b) (middle).
Instead of down-converting a narrow frequency photon and detecting
photons with anticorrelated frequencies, we prepare light with
anticorrelated frequencies and detect a narrow frequency band. The
CPI setup can be seen in Fig.~\ref{schematics}b)~(bottom) where a
pair of oppositely-chirped laser pulses enter into a
cross-correlator.  A narrow bandwidth of the output sum-frequency
generation (SFG) is detected on a standard photodiode as a function
of the time delay, $\Delta\tau$. We have shown that CPI can be used
in place of HOM interference to obtain the same benefits of
quantum-optical coherence tomography~\cite{Abouraddy2002a} with
dramatically larger signal and a straightforward means of control
over intrinsic signal artifacts~\cite{Lavoie2009a}.

Several quantum-interference effects are based on modifications of
the HOM interferometer, such as the three shown in  Figs.~1a)-c)
(upper).  In Fig.~1a) photon pairs are detected in one output. The
photon bunching leading to the HOM dip gives rise to
phase-insensitive constructive interference, a \emph{HOM peak}, in
the coincidence rate of these detectors~\cite{Mattle1996a}. In
Fig.~1b) bandpass filters centred at different wavelengths are
placed before the detectors. The coincidence rate in this device
exhibits phase-sensitive interference, but at a wavelength that
depends on the frequency difference of the filters~\cite{Ou1988a,
Legero2004a}. The wavelength of the interference, referred to as
\emph{quantum beating}, can be much longer than the wavelength of
the light. In Fig.~1c) the output of the HOM is fed into a
Mach-Zehnder interferometer. The output of the first interferometer
can be approximated by a two-photon \textsc{noon} state,
$|\psi\rangle \sim |2\rangle|0\rangle+|0\rangle|2\rangle$, which
exhibits \emph{phase super-resolution} (PSR), manifesting as a
wavelength of interference two times shorter than that of the light
passing through the interferometer.

For the experimental realization of classical analogues of these
three quantum effects, we use a modelocked ti:sapphire laser (centre
wavelength $790$~nm, pulse duration $100$~fs FWHM, average power
$2.8$~W, repetition rate $80\,\mbox{MHz}$) as the light source. The
beam is split at a 50:50 beam splitter. Its two outputs pass through
a grating-based stretcher~\cite{Martinez1988a} and
compressor~\cite{Treacy1969a} to generate chirped pulses
approximately $54\,\mbox{ps}$ long and anti-chirped pulses
$48\,\mbox{ps}$ long (FWHM), respectively.  The difference in pulse
durations is due to slightly different bandwidths rather than
different chirp rates. Because the stretcher and the compressor are
aligned to generate an equal but opposite chirp rate, at any given
time the sum of the instantaneous frequencies of the two pulses is
constant. For more details, see~\cite{Kaltenbaek2008a}.

Fig~\ref{schematics}b)(lower) shows the basic CPI setup. The
horizontally-polarized chirped and antichirped beams are combined at
a 50:50 beam splitter. From there the outputs travel along two
different spatial paths, one of which contains an adjustable path
delay. The polarization in one of the paths is rotated to vertical
by a half-wave plate. This allows the recombination of the beams
into a single spatial mode, but with orthogonal polarizations, at a
polarizing beam splitter.  That mode is focused onto a
$0.5\,\mbox{mm}$ thick type-II phase-matched $\beta$-Barium Borate
crystal for SFG. High-pass filters separate the SFG from the
fundamental signal. A narrow bandwidth of the SFG is filtered using
gratings and a slit and is detected using an amplified Si photodiode
(Thorlabs PDA36A).

\begin{figure}
\begin{center}
\includegraphics[width=1 \columnwidth]{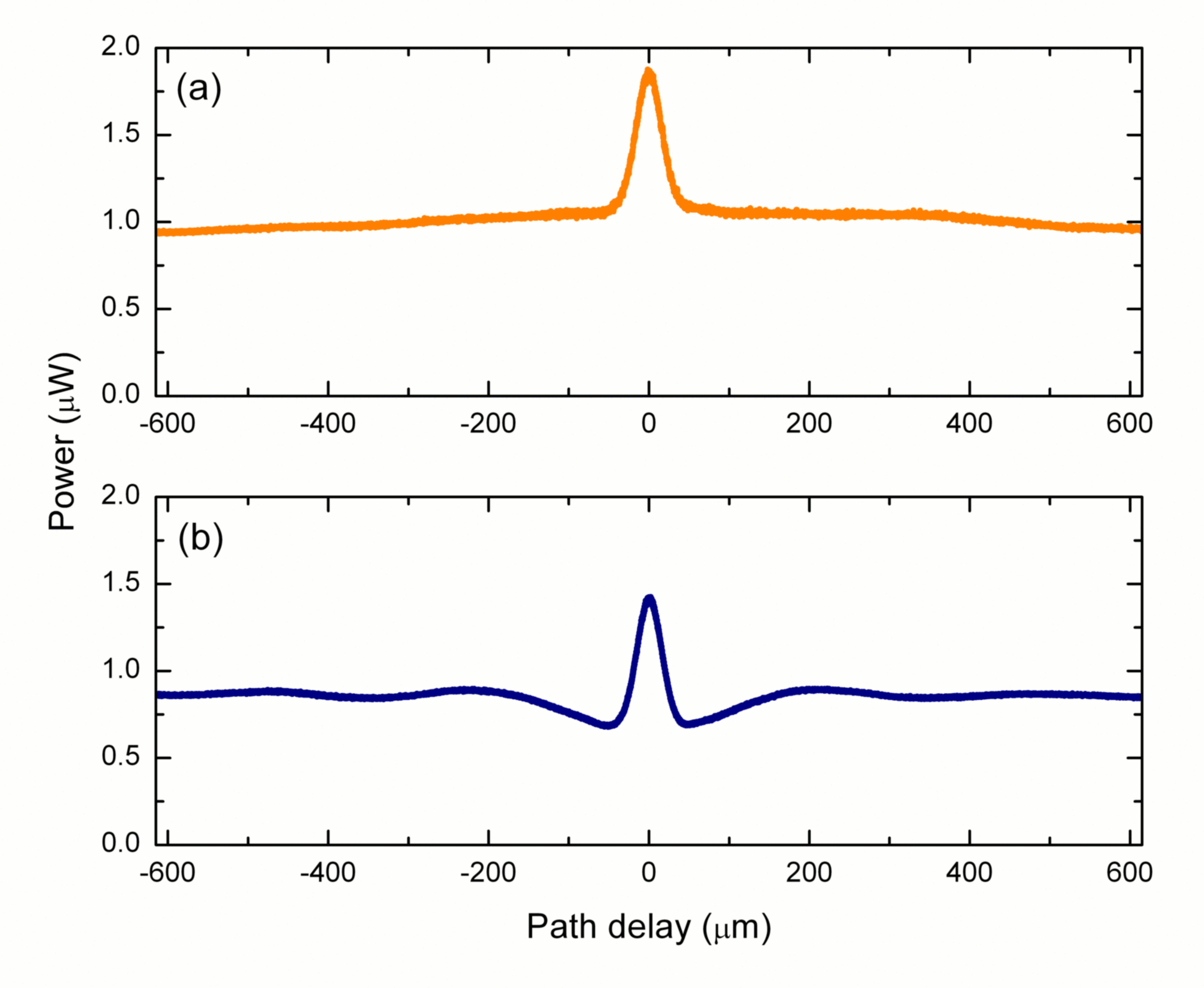}
\caption{Phase-insensitive constructive interference in CPI.  The
system was set up as depicted in Fig.~\ref{schematics}a)~(bottom) as
the time-reversed version of the two-photon interferometer in
Fig.~\ref{schematics}a)~(top). In b) the configuration was the same
as in a) except that we blocked $2.0\pm 0.3\,\mbox{nm}$ near the
centre wavelength of the chirped pulses in the stretcher. Both plots
show the measured photodiode signal as a function of delay. The data
in a) shows a phase-insensitive interference peak with visibility
$76\pm 2\,\%$ and FWHM $42\pm 2\,\mu\mbox{m}$. The data in b) shows a peak 
with similar visibility and width, but with two new features where the
signal drops at +/-$50\pm 4\,\mu\mbox{m}$ by $20\pm 1\,\%$ of the
plateau signal level.\label{HOMpeak}}
\end{center}
\end{figure}

The HOM peak can be observed in the quantum interferometer shown in
Fig.~\ref{schematics}a) (upper).  Time-reversing this setup requires
combining the oppositely-chirped laser pulses at a beamsplitter
before the input to the cross-correlator as shown in
Fig.~\ref{schematics}a) (middle). Fig.~\ref{HOMpeak}a) shows the
resulting interferogram as a function of delay. The path length was
varied in the delay arm of the cross-correlator by moving a motor
with a constant velocity of $0.500 \pm 0.005\,\mbox{mm}/\mbox{s}$.
Simultaneously, data was acquired with a sample rate of
$12\,\mbox{kHz}$. The gratings and the slit were adjusted to filter
the SFG with a bandwidth of $0.4\,\mbox{nm}$ FWHM around the center
wavelength $395.2 \pm 0.1\,\mbox{nm}$. The sole feature in the
interferogram is a phase-insensitive constructive interference peak
with visibility $76 \pm 2\,\%$. In Ref.~\cite{Resch2004a} it was
shown that an absorber that removes a narrow portion of the spectrum
near the centre frequency $\omega_0$ in front of the single-photon
detectors leads to a reduced coincidence rate close to the HOM peak.
This feature can be interpreted as enhanced absorption through a
photon exchange effect. In our experiment, we achieve an analogous
signal by blocking a $2.0 \pm 0.3\,\mbox{nm}$ band of the spectrum
at the centre frequency by placing an Allen key in the beam inside
the stretcher. Fig.~\ref{HOMpeak}b) is the resulting interferogram
clearly showing the appearance of two dips for delay positions just
outside the peak.

\begin{figure}
\begin{center}
\includegraphics[width=1 \columnwidth]{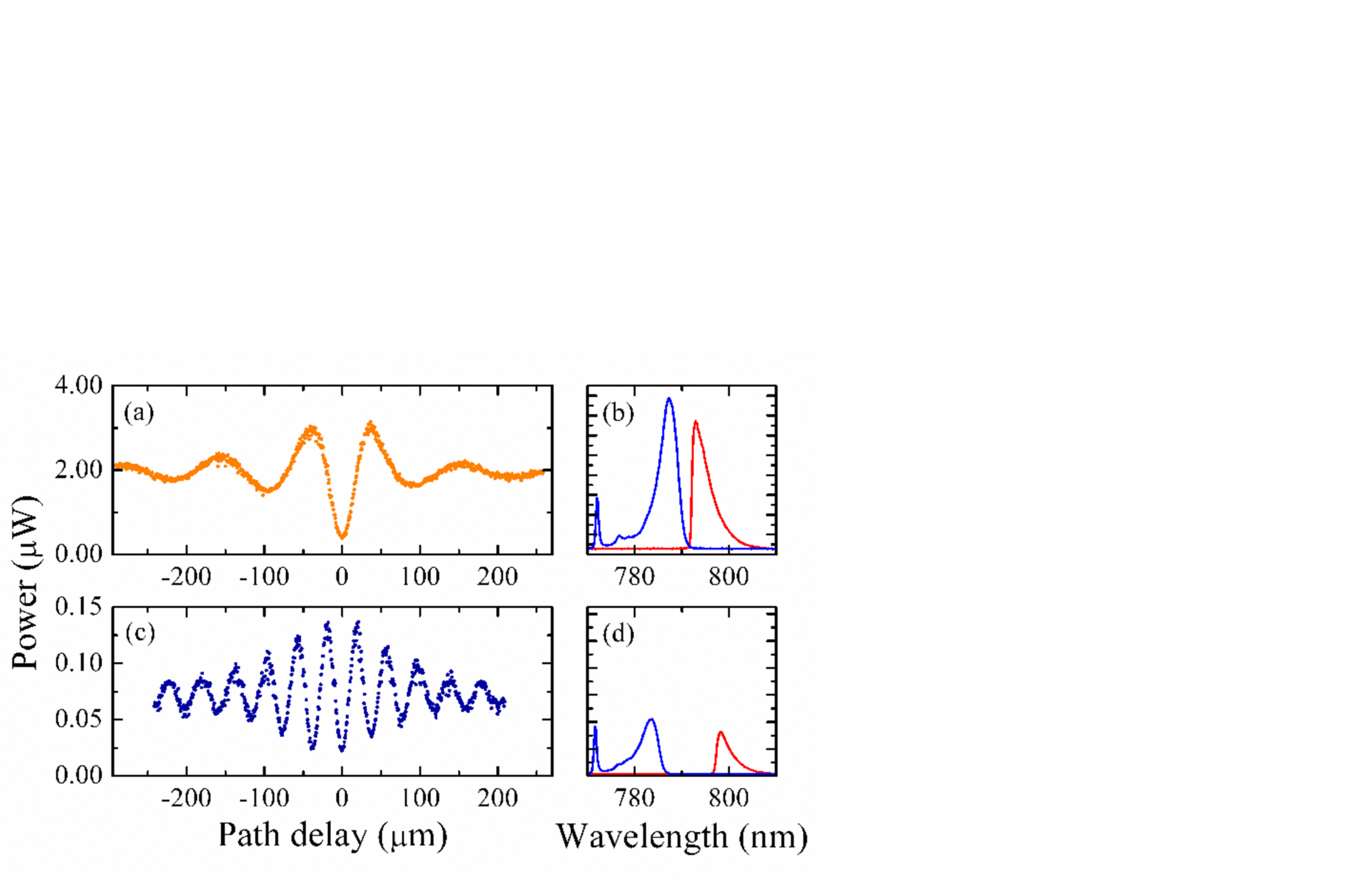}
\caption{``Quantum'' beating in CPI.  Filtering different spectral
components of the chirped and anti-chirped input plays the same role
in CPI as filtering different spectral components in the HOM
interferometer shown in Fig.~\ref{schematics}b)~(top).  Measured interference
patterns a) and c), and the corresponding spectra of the chirped
(red) and anti-chirped (blue) beams b) and d), respectively, are
shown.  The measured fringe spacing is a) $115\pm15\mu\mbox{m}$ and
c) $40\pm2\,\mu\mbox{m}$.  This is in good agreement with theory
where the fringe spacing is determined by the frequency difference
between the chirped and anti-chirped spectra.
\label{beating}}
\end{center}
\end{figure}

Quantum beating was originally observed in the quantum
interferometer shown in Fig.~\ref{schematics}b)(upper)
\cite{Ou1988a}. This is a standard HOM interferometer where
interference filters with different bandpasses are placed in front
of the detectors \emph{after} the interferometer. Time-reversing
this setup requires filtering different bandwidths of the chirped
and anti-chirped beams \emph{before} the interferometer.  We
inserted razor blades into the stretcher and compressor to block
spectral components of the light. The measured spectra for two
different positions of the razor blades are shown in
Figs.~\ref{beating}b) and d). We measured the SFG signal as a
function of delay by moving a motor in the delay path in steps of
$3\,\mu\mbox{m}$.  Note that in this configuration we used a stepper
motor and took data at discrete positions, whereas for the other
data we moved the motor and took data continuously. This accounts
for the qualitative difference between the appearance of these data
sets and the others. The SFG signal was detected within a bandwidth
of $0.3\pm 0.1\,\mbox{nm}$ FWHM around $394.5\pm0.1\,\mbox{nm}$.

The resulting CPI interferograms as functions of path delay are
shown in Figs.~\ref{beating}a) and c). Both signals clearly exhibit
interference fringes but with periods much larger than the
wavelength of the light. This is the same characteristic feature
that was oberved in the quantum beating
experiment~\cite{Ou1988a,Legero2004a}. For Figs.~\ref{beating}a) and
c) the difference frequency, as determined by the peaks of the
spectra, between chirp and anti-chirp is $17\pm1\,\mbox{ps}^{-1}$
and $45\pm1\,\mbox{ps}^{-1}$, respectively. From these difference
frequencies, we expect the corresponding fringe spacings to be
$111\pm7\,\mu\mbox{m}$ and $42\pm1\,\mu\mbox{m}$. Both are in good
agreement with the measured fringe spacings, $115\pm15\,\mu\mbox{m}$
and $40\pm2\,\mu\mbox{m}$, and much larger than either the
wavelength of the SFG, $0.395\,\mu\mbox{m}$, or the chirped pulses,
$0.790\,\mu\mbox{m}$.

Two-photon phase super-resolution can be observed in the
interferometer shown in Fig.~\ref{schematics}c) (upper).  The output
of a balanced HOM interferometer is fed into a Mach-Zehnder
interferometer.  HOM interference causes photon pairs to bunch
together, creating number-path entangled states. These exhibit
interference fringes, as measured in the coincidence rate of the
detectors, at half the classical period.  For the time-reversed
version, one could employ a spatial encoding as depicted in
Fig.~\ref{schematics}c)~(middle). However, we used an equivalent
transformation on the polarization degree of freedom by simply
inserting a half-wave plate, oriented at $22.5^\circ$, before the
SFG as shown in Fig.~\ref{schematics}c)~(bottom).

\begin{figure}
\centering
\includegraphics[width=1 \columnwidth]{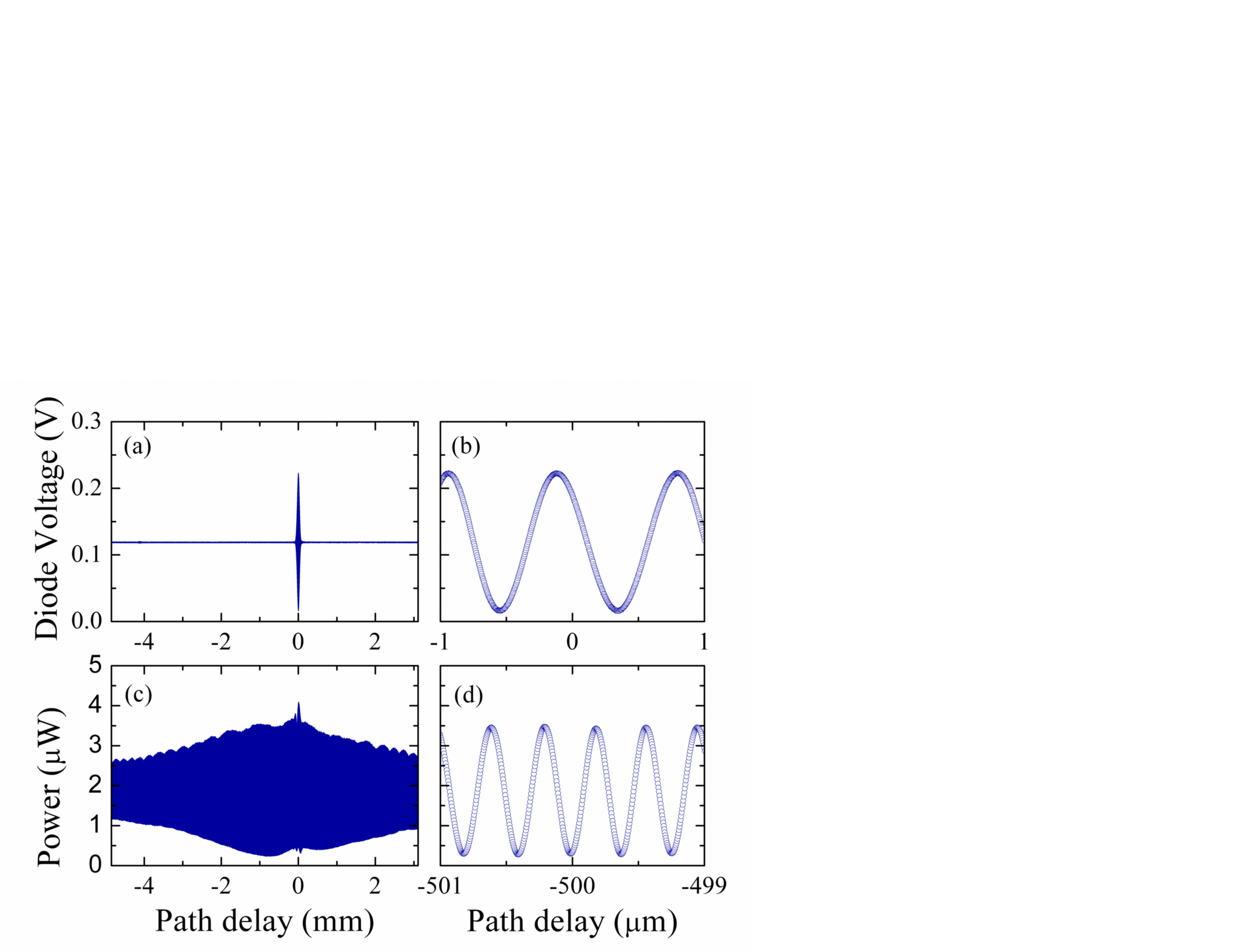}
\caption{White-light interference pattern and phase super-resolution
in CPI. a) \& b) shows the white-light interference pattern
generated by the chirped pulse.  c) \& d) show the SFG signal
detected in the modified CPI depicted in Fig.~1c) (lower).  By
comparing b) \& d) one can clearly see the reduction of the fringe
wavelength in CPI; this is phase super-resolution. In addition, by
comparing the signals in a) \& c), we see that the CPI coherence
length is roughly $80$ times longer than the white-light coherence
length. \label{NOON}}
\end{figure}

Figs.~\ref{NOON}c)~\&~d) show the results of a continuous scan of
the SFG signal over the path delay in the cross-correlator. For
comparison, Figs.~\ref{NOON}a)~\&~b) show a white-light
interferogram taken by replacing the half-wave plate with a
polarizer at $45^\circ$ and measuring the fundamental light with a
fast photodiode (Thorlabs DET100A). In both cases, the path delay
was continuously scanned by moving a motor with a velocity of
$0.500\pm0.005\,\mbox{mm}/\mbox{s}$ while the signal was recorded
with a sample rate of $250\,\mbox{kHz}$. The SFG signal was detected
within a bandwidth of $0.09\,\mbox{nm}$ FWHM around
$394.9\pm0.1\,\mbox{nm}$. The entire data set took $7\,\mbox{s}$ to
accumulate with a resolution of about $100$ points per fringe. The
fringes in Figs.~\ref{NOON}b)~\&~d) for white-light and CPI have
$87.1\pm0.2\,\%$ and $84.5\pm0.5\,\%$ visibility, respectively.

One can clearly see that the CPI fringe period,
$395\pm4\,\mbox{nm}$, is half that of the white light,
$795\pm8\,\mbox{nm}$, demonstrating phase super-resolution. The PSR
signal in Fig.~\ref{NOON}d) is centered around $-500\mu\mbox{m}$ to
show it free of residual white-light interference due to imperfect
alignment. Comparing Figs.~\ref{NOON}a)~\&~c) we see another
characteristic in our classical system often associated with quantum
interference. The coherence length for the white-light interference
pattern is $63.5\pm0.3\mu\mbox{m}$ FWHM, in good agreement with
expectations from the bandwidth of the chirped pulse of
$10\,\mbox{nm}$ FWHM at $790\,\mbox{nm}$. The width of the PSR
interferogram, on the other hand, is approximately $5\mbox{mm}$
FWHM, a factor of $80$ larger. Under ideal conditions, perfect mode
overlap, matching chirp rates and bandwidths, and assuming Gaussian
spectra, the width of the interferogram is calculated to be
$19\,\mbox{mm}$, the length of the chirped pulses.

Phase super-resolution has previously been shown in a multiport
classical interferometer in the coincidence rate between a set of
photon counters \cite{Resch2007a}.  The CPI approach demonstrated
here is different in two important ways. It does not rely on
single-photon detection facilitating rapid data accumulation.
Furthermore, it is the first observation of a classical analogue to
dramatically different one-photon and two-photon coherence lengths
that have been reported in entangled quantum systems
~\cite{Fonseca1999a,Edamatsu2002a}.

We have shown classical analogues to three archetypical quantum
interference effects by making modifications to chirped-pulse
interferometry.  This work demonstrates the first observation of
classical analogue of the Hong-Ou-Mandel peak and quantum beating.
We have also demonstrated a new method for observing phase
super-resolution in a classical interferometer suitable for rapid
data acquisition and exhibiting a coherence length much longer than
that of the laser light.  These results constitute a step toward answering
a central question in quantum information science as to which
phenomena require quantum resources and which can be achieved
classically.

We thank Devon Biggerstaff and Gregor Weihs for valuable
discussions. This work was supported by NSERC, OCE, and CFI; R.K.
acknowledges financial support from IQC. J. L. acknowledges
financial support from the Bell Family Fund.

\end{document}